\documentclass[10pt,twocolumn,pra,aps,showkeys,showpacs,superscriptaddress,longbibliography]{revtex4-1}

\usepackage{amsmath}
\usepackage{amssymb}
\usepackage{graphicx}
\usepackage[usenames,dvipsnames,svgnames,table]{xcolor}
\usepackage[unicode=true,bookmarks=false,breaklinks=false,pdfborder={0 0 1},colorlinks=true]{hyperref}
\hypersetup{citecolor=magenta,urlcolor=blue}

\usepackage{bm}
\usepackage{braket}
\newcommand{\inter}{\mathrm{int}}

\begin{document}

\title{Semi-synthetic zigzag optical lattice for ultracold bosons}

\author{E.~Anisimovas}
\email{egidijus.anisimovas@ff.vu.lt}
\affiliation{Institute of Theoretical Physics and Astronomy, Vilnius University, 
Saul\.{e}tekio 3, LT-10222 Vilnius, Lithuania}

\author{M.~Ra\v{c}i\={u}nas}
\affiliation{Institute of Theoretical Physics and Astronomy, Vilnius University, 
Saul\.{e}tekio 3, LT-10222 Vilnius, Lithuania}

\author{C.~Str\"{a}ter}
\email{cstraeter@pks.mpg.de}
\affiliation{Max-Planck-Institut f\"{u}r Physik komplexer Systeme, 
N\"{o}thnitzer Stra{\ss}e 38, D-01187 Dresden, Germany}

\author{A.~Eckardt}
\email{eckardt@pks.mpg.de}
\affiliation{Max-Planck-Institut f\"{u}r Physik komplexer Systeme, 
N\"{o}thnitzer Stra{\ss}e 38, D-01187 Dresden, Germany}

\author{I.~B.~Spielman}
\email{ian.spielman@nist.gov}
\affiliation{Joint Quantum Institute, University of Maryland, College Park, Maryland 20742-4111, USA}
\affiliation{National Institute of Standards and Technology, Gaithersburg, Maryland 20899, USA}

\author{G.~Juzeli\={u}nas}
\email{gediminas.juzeliunas@tfai.vu.lt}
\affiliation{Institute of Theoretical Physics and Astronomy, Vilnius University, 
Saul\.{e}tekio 3, LT-10222 Vilnius, Lithuania}

\date{\today}

\begin{abstract}
We propose a cold-atom realization of a zigzag ladder. The two legs of the ladder 
correspond to a ``synthetic'' dimension given by two internal (spin) states of the 
atoms, so that tunneling between them can be realized as a laser-assisted process. 
The zigzag geometry is achieved by employing a spin-dependent optical lattice with the 
site position depending on the internal atomic state, i.~e.\ on the ladder's leg.
The lattice offers a possibility to tune the single-particle dispersion
from a double-well to a single-minimum configuration. In contrast to previously considered 
semi-synthetic lattices with a square geometry, the tunneling in the synthetic dimension 
is accompanied by spatial displacements of atoms. Therefore, the atom-atom interactions 
are nonlocal and act along the diagonal (semi-synthetic) direction. We investigate the 
ground-state properties of the system for the case of strongly interacting bosons. 
In particular, we find that the interplay between the frustration induced by the magnetic 
field and the interactions gives rise to an interesting gapped phase at fractional 
filling factors corresponding to one particle per magnetic unit cell.
\end{abstract}

\maketitle

\section{Introduction}

Optical lattices provide a unique tool for simulating quantum condensed matter physics
using ultracold atoms \cite{Lewenstein2007,Bloch2008a,Lewenstein2012}. These lattices 
can be enriched by introducing laser-coupled internal atomic states 
\cite{Javanainen2003,Jaksch2003,Osterloh2005,Dalibard2011,Goldman2014,goldman16review}
that can play the role of an extra ``synthetic'' dimension \cite{Boada12PRL,Celi14,Price15PRL}. 
For example, a semi-synthetic square lattice results from the combination of the interlayer 
tunneling among the sites of a one-dimensional optical lattice and laser-assisted transitions 
between the onsite atomic levels. If the laser coupling is accompanied by a recoil in the 
lattice direction, the semi-synthetic lattice acquires a uniform magnetic flux traversing 
the square plaquettes \cite{Celi14}. This leads to the formation of chiral edge states 
in the resulting quantum Hall ribbon 
\cite{Celi14,Fallani16Science,Spielman16Science,Fallani16arXiv,An2016}. 
A characteristic feature of the square geometry is that the atom-atom interaction 
is long-ranged in the synthetic dimension but short-ranged in the real dimension 
\cite{Celi14,Zhai15PRL,Shenoy15PRA,Cooper15PRAR}. 

In this work, we depart from the square geometry and find the ground states of a semi-synthetic 
optical \emph{zigzag} lattice which can be created combining a spin-dependent one-dimensional 
optical lattice with laser-induced transitions between the atomic internal states 
\cite{[{On a single-particle level, a possibility of creating a non-square semi-synthetic 
geometry was recently considered by }] [{. The proposal relies on experimentally more 
challenging additional diagonal tunnelings between the original sites of a semi-synthetic 
square lattice.}] Suszalski16PRA}.
The lattice is affected by a tunable homogeneous magnetic flux, and furthermore features 
nonlocal interactions along the semi-synthetic directions that connect different internal
states situated at different spatial locations, see also Ref.~\onlinecite{Chhajlany16}.
Generation of magnetic fluxes in an effectively one-dimensional setting is intriguing and
was recently considered in Ref.~\onlinecite{Grass15}. Nonlocal interactions are also an 
important goal in recent experiments, and such interactions have been engineered via 
superexchange \cite{Greif2015af,Hulet2015af,Boll2016af,Cheuk2016af} 
dipole-dipole coupling \cite{Lahaye2009review,Yan2013polarmolecules,Frisch2015magdip,Baier2016bhm},
or Rydberg dressing \cite{Pupillo2010rydberg,Viteau2011rydberg,Glaetzle2015rydberg,Labuhn2016}.
We investigate the ground-state properties of the proposed system for the case 
of bosonic atoms with strong interactions using the density-matrix renormalization group 
\cite{Vidal04,Schollwock2011,orus14review} calculations. We find that the interplay between 
the frustration induced by the magnetic flux and the interactions gives rise to an 
interesting gapped phase at fractional per-site filling fractions corresponding to
one particle per magnetic unit cell. 

The paper has the following structure. The single-particle model is formulated in 
Sec.~\ref{sec:2a} introducing the experimentally motivated lattice setup. The model is solved 
and analyzed in Sec.~\ref{sec:2b}-\ref{sec:2d}, in particular, in Sec.~\ref{sec:2d} we 
explore the manifestation of the resulting band structure via Bloch oscillations of a wave 
packet in a tilted lattice. Section \ref{sec:manybody} is devoted to the many-body phases 
supported by the semi-synthetic zigzag lattice. The concluding Sec.~\ref{sec:summary}
summarizes the findings.

\section{Single-particle Hamiltonian}

\subsection{Lattice setup}\label{sec:2a}

\begin{figure}
\includegraphics[width=84mm]{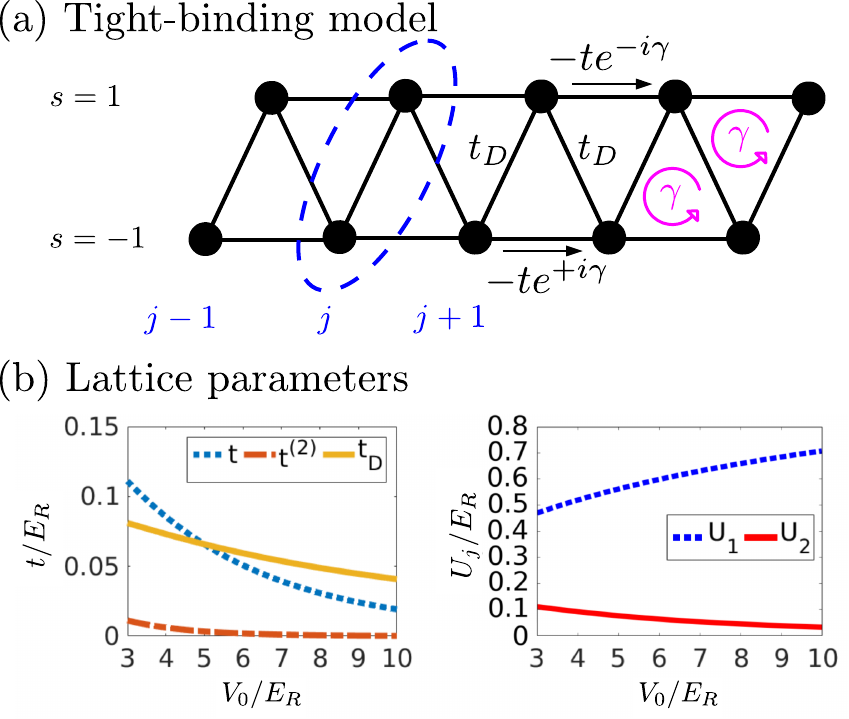} 
\caption{\label{fig:setup} 
(a) The semi-synthetic zigzag lattice corresponding to the tight-binding Hamiltonian 
(\ref{eq:ham_tight}). The lattice is affected by a non-staggered flux 
$\gamma = a\tilde{\kappa}/2 = \pi\tilde{\kappa}/\kappa$ over triangular plaquettes.
(b) Tight-binding parameters. Left: Horizontal nearest neighbor ($t$) and next-nearest 
neighbor ($t^{(2)}$) hopping parameters together with diagonal tunneling strength $t_{D}$ 
as a function of the scaled lattice depth $V_{0}/E_{R}$ for $\Omega = 0.2020E_{R}$; 
Right: Normalized interactions $U_{1}$ and $U_{2}$ as a function of the scaled lattice depth.}
\end{figure}

We consider bosonic atoms with two relevant internal states labeled with the (quasi-)spin 
index $s = \pm 1$. To create the semi-synthetic zigzag lattice shown in 
Fig.~\ref{fig:setup}(a), the atoms are confined in a one-dimensional periodic trapping 
potential $V \propto \pm \cos(\kappa x)$, opposite for each internal state. In addition, the 
two quasi-spin states are coupled by laser-induced transitions characterized by a Rabi 
frequency $\Omega$ and a recoil wave vector $\tilde{\kappa}\mathbf{e}_{x}$ aligned along 
the lattice direction $\mathbf{e}_x$. The resulting single-particle Hamiltonian is 
\begin{equation}
\label{eq:ham_exact}
  H = \frac{\hat{p}^{2}}{2m} 
  + \frac{V_{0}}{2} \cos(\kappa x)\,\sigma_{z} 
  + \hbar\Omega\left(\sigma_{+}e^{i\tilde{\kappa}x}
  + \sigma_{-}e^{-i\tilde{\kappa}x}\right),
\end{equation}
where $V_{0}$ is the height of the trapping potential while $\sigma_{z}$ and
$\sigma_{\pm} = \sigma_{x} \pm i\sigma_{y}$ denote the standard Pauli spin matrices and 
combinations thereof. 

\begin{figure}
\includegraphics[width=84mm]{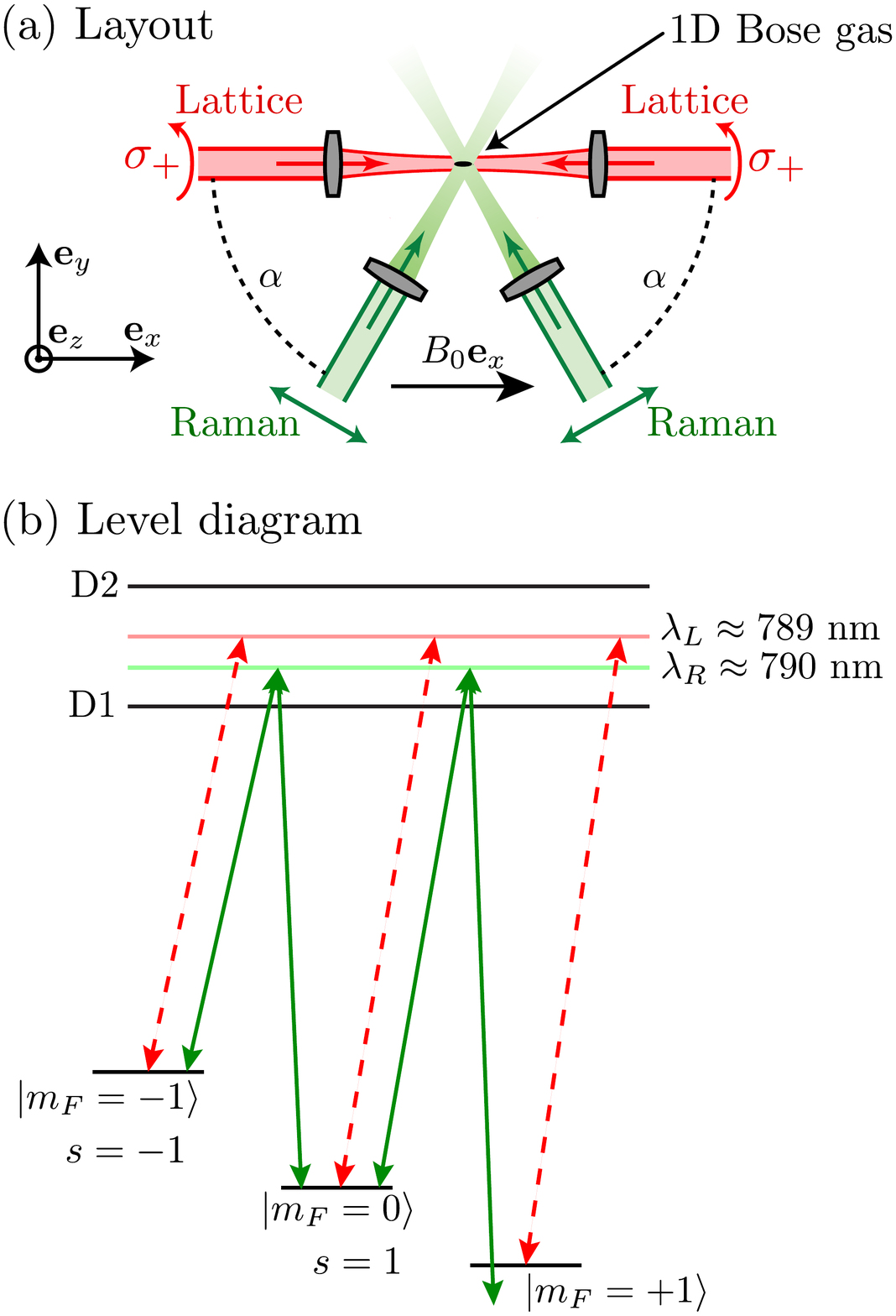} 
\caption{\label{fig:experimental} 
(a) Schematic layout of proposed experimental setup. 
A bias magnetic field ${\bf B}_0 = B_{0}{\bf e}_{x}$ Zeeman-splits the hyperfine spin states 
of $^{87}{\rm Rb}$ atoms. A counter-propagating pair of $\sigma_{+}$ polarized laser beams 
(shown in red) with $\lambda_{L} = 4\pi / \kappa \approx789\ {\rm nm}$ form a spin-dependent 
lattice with opposite signs for atoms in the $m_F=-1$ and $m_F=0$ states
\cite{Deutsch1998,McKay2010,Goldman2014}. This traps the $\ket{F = 1,m_{F} = -1, 0}$ 
states on lattice sites shifted by half the lattice constant $a/2=\pi/\kappa$ providing 
a semi-synthetic zigzag lattice. Two horizontally polarized lasers (shown in red) at 
$\lambda_{R}=2\pi/\tilde{\kappa}\approx790\ {\rm nm}$ resonantly couple the two spin 
states producing the flux $\gamma =  \tilde{\kappa}a/2=\pi/2$ tuned by taking the angle 
$\alpha\approx60^{\circ}$ between the laser beams.  
(b) Level diagram. The $\sigma_+$ polarized lattice laser beams (dashed red arrows)
shift individual $m_F$ states, but do not drive transitions. 
The strength of the state-dependent contribution to this shift is maximized when the beams 
propagate parallel to ${\bf B}_0$.
The Raman lasers (solid green arrows) are tuned to be in resonance with $m_F = -1 \rightarrow 0$ 
transition, but detuned from the $m_F=0\rightarrow +1$ transition.
Note that the Zeeman splitting of the ground state $F = 1$ hyperfine manifold is shown 
on a greatly exaggerated scale.}
\end{figure}

The out-of-phase optical lattice can be produced by taking the quasi-spin states with
$s = \pm 1$ to be the ground state $^1$S$_0$ and the 
long-lived excited state $^{3}$P$_{0}$ of the alkaline-earth(-like) atoms, such as Ytterbium 
\cite{Fallani16arXiv} or Strontium \cite{Kolkowitz2016socSr,Wall2016socSr}, for which the 
excited state has a typical lifetime far exceeding the experimental time scale
\cite{Gerbier2010,Dalibard2011,Fallani16arXiv,Kolkowitz2016socSr}. In contrast to the recent 
experiments \cite{Fallani16arXiv,Kolkowitz2016socSr}, the atoms are to be trapped at an 
anti-magic (rather than magic) wavelength to have the opposite trapping potentials for the 
two atomic internal states. 
Alternatively, one may use two Raman-coupled hyperfine atomic states 
$\left| F, m_{F} \right\rangle$ with projections $m_{F} = 0$ and $m_{F} = -1$ from the $F = 1$ 
ground-state manifold of the $^{87}{\rm Rb}$ atoms \cite{Lin2011}
as the two quasi-spin states (see Fig.~\ref{fig:experimental}). 
The lattice potential $V\propto\pm\cos(\kappa z)$ is then obtained 
by balancing the vector and scalar light shifts of a state-dependent lattice 
\cite{Deutsch1998,Goldman2014}. This can be done by using a standing wave of a circularly 
(either $\sigma_{+}$ or $\sigma_{-}$) polarized light, and detuning slightly away from the 
frequency at which the scalar light shift is exactly zero.

\subsection{Tight binding approximation}\label{sec:2b}

We focus on a sufficiently deep lattice potential with the depth 
$V_{0}$ typically exceeding the recoil energy $E_{R} = \hbar^{2}\kappa^{2}/8m$ five times. 
In this regime, a tight-binding approach is appropriate. We use the index $j$ to label
the sites along the physical (long) direction, and the internal states with $s = \pm 1$ 
are interpreted as sites along the synthetic dimension \cite{Celi14}. This provides
a semi-synthetic zigzag lattice depicted in Fig.~\ref{fig:setup}(a). 

To proceed with the tight-binding approach, we introduce the Wannier functions $w_j (x)$ 
for the atomic motion in the one-dimensional cosine potential $V(x) = V_{0}\cos(\kappa x)/2$ 
oscillating with the spatial periodicity $a = 2\pi/\kappa$. The functions 
$w_j (x) \equiv w_0 (x - ja)$ 
are localized at the potential minima $x_j = aj$. The Wannier basis for the two spin 
states with $s = \pm 1$ is thus given by 
\begin{equation}
\label{eq:wannier}
  w_{s,j}(x) = w_{0}(x - sa/4 - ja),
\end{equation}
where  for convenience the origin of the $x$ axis has been shifted to the midpoint between 
the neighboring $s = \pm 1$ sites. The locations of the opposite spin states differ by 
$a/2$, i.~e.\ by a half of the lattice constant. 

Matrix elements for tunneling along the real dimension have the usual form 
\begin{equation}
\label{eq:t}
 -t = \int w_{s,j+1}^{*}(x) \left[ \frac{\hat{p}^{2}}{2m}
 - \frac{V_{0}}{2} \cos(\kappa x) \right] w_{s,j}(x)\,dx.
\end{equation}
With the minus sign absorbed into the definition in Eq.~(\ref{eq:t}), the quantity $t$ 
is real and positive. Matrix elements for the laser-assisted tunneling along the two 
``diagonal'' directions of the semi-synthetic lattice are obtained by overlapping the Wannier 
functions weighted with the position-dependent laser coupling term:
\begin{subequations}
\label{eq:td}
\begin{equation}
  \int w_{1,j}^{*}(x) \,\Omega \, e^{i\tilde{\kappa}x} w_{-1,j}(x) \, dx
  = t_{D} \, e^{i\tilde{\kappa}aj},
\end{equation}
and 
\begin{equation}
  \int w_{1,j+1}^{*}(x) \,\Omega \, e^{i\tilde{\kappa}x} w_{-1,j}(x)\, dx
  = t_{D} \, e^{i\tilde{\kappa}a\left(j+1/2\right)}.
\end{equation}
\end{subequations}
Here the amplitude $t_{D}$ is determined by both the Rabi frequency $\Omega$ and the 
overlap integral $\rho$ between the neighboring Wannier functions for the opposite 
spin states:
\begin{equation}
\label{eq:td_rho}
  t_{D} = \Omega\rho, \quad
  \rho = \int w_{0}^{*}(x-a/4) \, e^{i\tilde{\kappa}x} \, w_{0}(x+a/4)\,dx.
\end{equation}

Within the tight-binding approach, we introduce the Bose operators $c_{s,j}$ and 
$c_{s,j}^{\dagger}$ to describe the annihilation and creation of atoms on the sites 
$(s,j)$ of the semi-synthetic zigzag lattice. By adding appropriate phase factors to these 
operators $c_{s,j}\rightarrow c_{s,j} \, e^{-ijsa\tilde{\kappa}/2}$, one arrives 
at the tight-binding Hamiltonian with complex-valued tunneling elements 
$e^{\pm isa\tilde{\kappa}/2}$ along the long direction (real dimension) and real-valued 
tunneling $t_{D}$ along the diagonal semi-synthetic directions:
\begin{equation}
\label{eq:ham_tight}
\begin{split}
  H &= t_{D} \sum_{j} \left[ c_{1,j}^{\dag} c_{-1,j}
    + c_{1,j-1}^{\dag} c_{-1,j} \right] \\
   &- t \sum_{j,s} c_{s,j+1}^{\dag} c_{s,j} \, 
   e^{-isa\tilde{\kappa}/2} + \mathrm{H.c.}.
\end{split}
\end{equation}
Here the first contribution describes the diagonal (spin-flip) tunneling in the semi-synthetic 
lattice. The lattice is affected by a non-staggered flux 
$\gamma = a\tilde{\kappa}/2 = \pi\tilde{\kappa}/\kappa$ over triangular plaquettes due 
to the recoil, as illustrated in Fig.~\ref{fig:setup}(a).

Figure \ref{fig:setup}(b) displays the dependence of the tunneling parameters $t$ 
and $t_{D}$ on the lattice depth for the characteristic value of the Rabi frequency 
$\Omega = 0.2020\,E_{R}$. This particular choice of the laser strength leads to equal 
values of the two hopping parameters $t = t_{D}$ for the lattice depth $V_{0} = 5E_{R}$ 
subsequently used in the many-body calculations. Note that the ratio $t_{D}/t$ is tunable 
and increases linearly with the Rabi frequency $\Omega$. Couplings between more distant 
sites are much smaller and can be safely neglected.

\subsection{Single particle spectrum}
\label{sec:2c}

In terms of the momentum-space bosonic operators $\hat{c}_{s,k}^{(\dagger)}$ 
the Hamiltonian reads 
\begin{equation}
\label{eq:ham_spec}
  H = \sum_{k} 
  \begin{pmatrix} \hat{c}_{1,k}^{\dagger} & \hat{c}_{-1,k}^{\dagger} \end{pmatrix}
  \begin{pmatrix} h_{11} & h_{12} \\ h_{21} & h_{22} \end{pmatrix}
  \begin{pmatrix} \hat{c}_{1,k} \\ \hat{c}_{-1,k} \end{pmatrix},
\end{equation}
where $h_{12} = h_{21} = 2 t_{D} \cos(ka/2)$ and $h_{jj} = -2 t \cos [ka + \gamma(-1)^j]$, 
with the row index $j = 1,\,2$. To develop more intuition into the single-particle 
properties of the model, let us look at the case where $\tilde{\kappa} = \kappa / 2$. 
The flux over a triangular plaquette is then $\gamma = \pi/2$, so that the time-reversal 
symmetry is broken in the semi-synthetic lattice even though the flux over a full 
elementary cell evaluates to $2\gamma = \pi$. In passing we note that the 
time-reversal symmetry is preserved if the triangular plaquette of the zigzag lattice is 
pierced by a $\pi$ flux \cite{Santos2013PRA}. Returning to the situation where 
$\gamma = \pi / 2$, the two dispersion branches read 
\begin{equation}
  \label{eq:disp}
  \epsilon_{\pm}(k) = \pm 2 \sqrt{t^{2} \sin^{2}(ka) + t_{D}^{2} \cos^{2}(ka/2)}.
\end{equation}

\begin{figure}
\includegraphics[width=84mm]{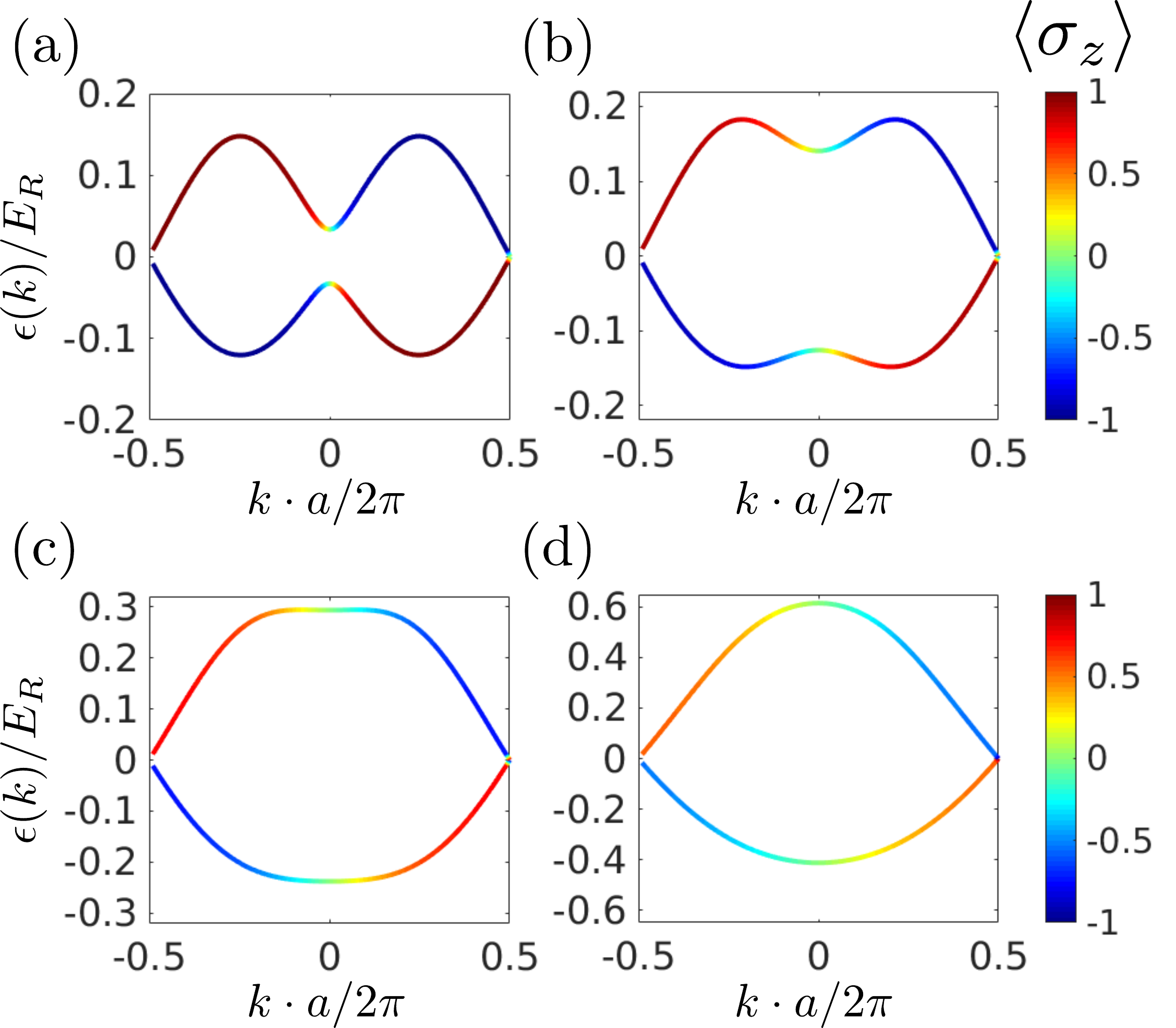} 
\caption{\label{fig:dispersion} Exactly calculated dispersion curves for $V_{0} = 5\,E_{R}$, 
$\kappa = 2 \tilde{\kappa}$ ($\gamma = \pi/2$) and various strengths of the spin-flip 
coupling: (a) $\Omega = 0.05$, (b) $\Omega = 0.2020$, (c) $\Omega = 0.4039E_{R}$ 
(the critical value), and (d) $\Omega = 0.8 E_{R}$. This corresponds to: 
(a) $t_{D} = 0.25 t$, (b) $t_{D} = t$, (c) $t_{D} = 2t$ (quartic dispersion at $k = 0$), 
and (d) $t_{D}=4t$.}
\end{figure}

The tight-binding dispersion (\ref{eq:disp}) is in a good agreement with the exact 
band structure shown in Fig.~\ref{fig:dispersion} for the zigzag lattice with 
$\tilde{\kappa} = \kappa / 2$ and $V_{0} = 5E_{R}$ corresponding to 
$\gamma=\pi/2$, $t = 0.0658E_{R}$ and $\rho = 0.3258$, with different values of 
$\Omega$ determining $t_{D} = \Omega\rho$. It is noteworthy that the dispersion becomes 
quartic around $k = 0$ for $t_{D} = t_{D,\mathrm{critical}} = 2t$ which corresponds 
to the critical Rabi frequency $\Omega_{\mathrm{critical}} = 2t / \rho$. For the lattice 
depth $V_{0} = 5E_{R}$ the critical Rabi frequency is 
$\Omega_{\mathrm{critical}} = 0.4039\,E_{R}$, and the resulting band structure is shown 
in Fig.~\ref{fig:dispersion}(c). Below the critical value, $t_{D} < 2t$, there are two symmetric 
minima at $ka = \pm \arccos[(t_{D}/2t)^{2}]$. Above the critical value, $t_{D} > 2t$, 
there is a single minimum at $k = 0$. 

We stress that the plots in Fig.~\ref{fig:dispersion}
represent the exact calculations which agree well with the tight-binding model for 
$\Omega$ up to the critical value $\Omega_{\mathrm{critical}}$ and a little above it. 
Yet for $\Omega = 0.8E_{R}\approx 2 \Omega_{\mathrm{critical}}$ (i.e. for 
$\Omega \rho \approx 4 t$), there is already a marked deviation from the 
tight-binding model due to mixing with higher orbital bands. In fact,
since the gap between the first and the second orbital bands is of the order of $2E_{R}$ at 
$V_{0}=5E_{R}$, the inter-band coupling becomes relevant only for larger $\Omega$ which 
is comparable to $E_{R}$, such as for $\Omega=0.8E_{R}$. 
This is approximately the regime where Zhou and Cui \cite{Zhou15PRBR} also saw deviations 
from the tight-binding model for a square semi-synthetic lattice.

The spin magnetization $\langle \sigma_z \rangle$ of the eigenstates is indicated by color 
in Fig.~\ref{fig:dispersion}. The red and blue colors correspond to a fully magnetized state 
with $s = 1$ and $s = -1$, respectively. In the case of weak coupling (upper panels) 
the dispersion has a double-well shape with a clear spin separation in different minima. 
For stronger coupling the spin states get increasingly mixed. At the critical value 
$\Omega=0.4039E_{R}$, the double well transforms to a single-minimum shape with a strong 
spin mixture. 

\subsection{Bloch oscillations}\label{sec:2d}

A characteristic feature of the zigzag lattice is the crossing of the two energy bands at 
the edges of the Brillouin zone $ka = \pm\pi$. In Fig.~\ref{fig:dispersion} we see that 
there is no band gap at these points and the spin polarization is preserved when moving 
from one energy band to the other at the Brillouin zone boundary
$ka = \pm\pi$. This is also true for other values 
of the flux $\gamma$. The absence of the gap is a consequence of the invariance of the 
Hamiltonian (\ref{eq:ham_exact}) under the spatial translation by half the lattice period 
$a/2$ followed by time reversal, the latter representing a spin flip combined with an 
inversion of the Peierls phase $\gamma \to -\gamma$ \footnote{In this discussion we are 
employing a usual definition of the time reversal symmetry involving the complex conjugation 
and spin reversal. On the other hand, in the tight binding picture the spin states are 
treated as sites in an extra dimension making a semi-synthetic zigzag lattice. Adopting 
such a point of view the time-reversal symmetry no longer involves the spin flip and thus 
has another meaning. This kind of the 'time-reversal symmetry' is implied in the paragraph 
following Eq.~(\ref{eq:ham_spec}).}. As a result, the period of Bloch oscillations is 
doubled, cf.\ Ref.~\onlinecite{Khomeriki16lzb}. 

\begin{figure*}
\includegraphics[width=126mm]{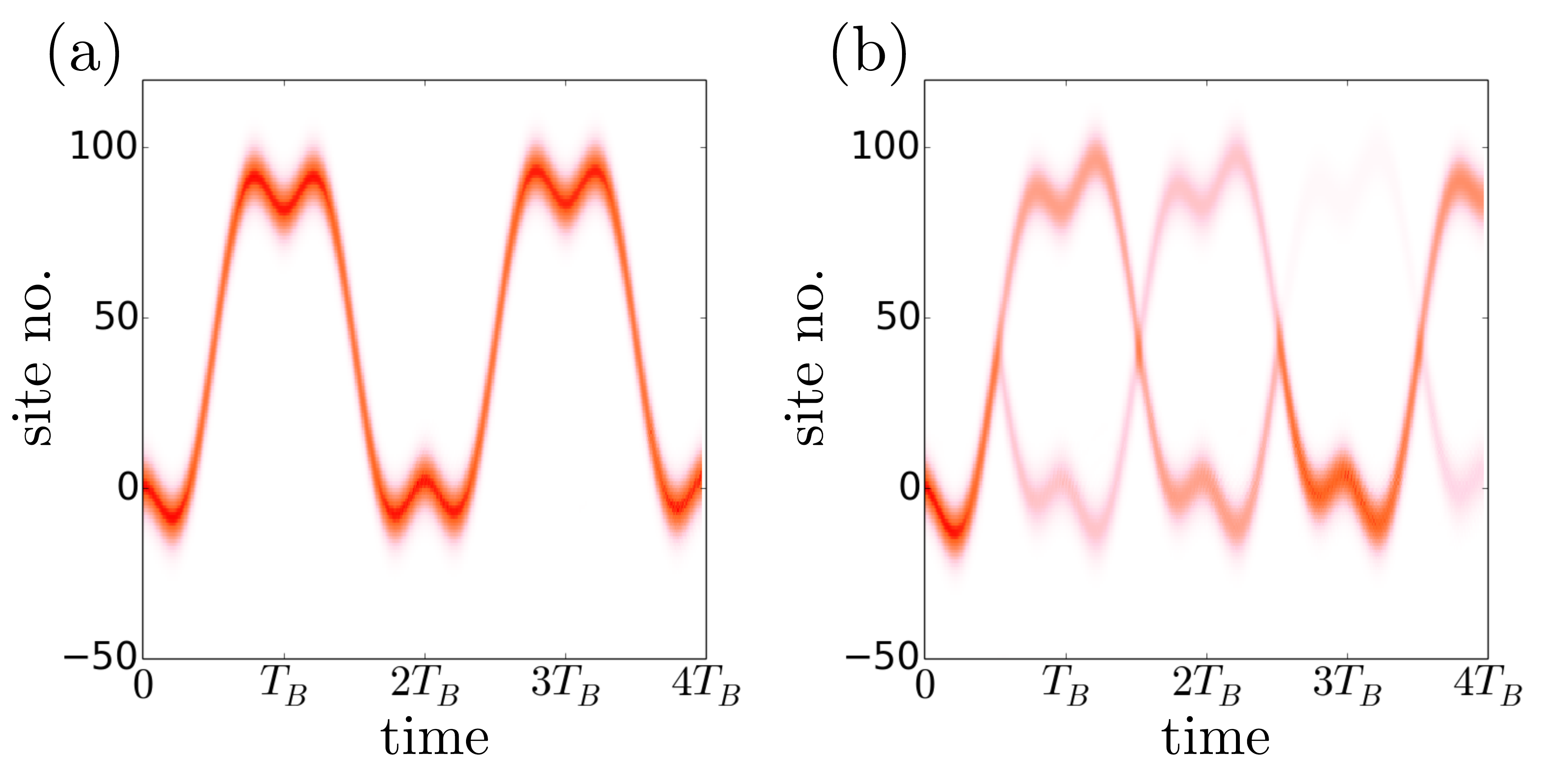} 
\caption{\label{fig:bloch} Bloch oscillations for a Gaussian wave packet in a tilted zigzag 
lattice. Panel (a) corresponds to the band structure presented in Fig.~2(b) of the
main text. In panel (b) the legs of the ladder legs are additionally biased by 
introducing a spin-dependent onsite energy shift $\pm 0.3\,t \sigma_z$, and the band gaps 
opened close to the Brillouin zone boundary lead to the Landau-Zener tunneling between the 
two bands.}
\end{figure*}

To illustrate the observable consequences of symmetry-related doubling of the Brillouin zone,
we performed a numerical simulation of a wave packet in the zigzag lattice. We prepared a 
Gaussian wave packet composed entirely of the states from the lower energy band close to 
$k = 0$ and initially situated at a certain position (referred to as site $j = 0$) in the 
real space. Under the influence of a lattice tilt the wave packet is scanning the single-particle
band structure while transferring diabatically between the two energy bands at the edges of the 
Brillouin zone. The results of our numerical simulation are shown in Fig.~\ref{fig:bloch}(a)
for the specific choice $t_D = t$, and clearly indicate the doubling of the Bloch period. 
To further clarify this effect, we contrast these results to those shown 
in Fig.~\ref{fig:bloch}(b) in which the onsite energies are modified by an additional 
spin-dependent bias $0.3\,t\sigma_{z}$. In such a situation the single-particle bands acquire small gaps 
at the Brillouin zone boundaries. As a consequence, the wave packet is split with the atoms 
being partially transferred into the other band each time the Brillouin zone boundary is 
reached.

\section{ Many-body effects}\label{sec:manybody}

\subsection{Interaction Hamiltonian}

To take interactions into account, the tight binding Hamiltonian (\ref{eq:ham_spec}) 
is complemented with the interaction term 
\begin{equation}
\label{eq:ham_int}
\begin{split}
  H_{\inter} &= \frac{U_{1}}{2} \sum_{j,s} n_{s,j} (n_{s,j}-1) \\
    &+ U_{2} \sum_{j} \left[ n_{1,j}  + n_{1,j-1} \right] n_{-1,j},
\end{split}
\end{equation}
where 
\begin{equation}
\label{eq:U1}
  U_{1} = U_{0} \int \left|w_{0}(x) \right|^{4}\,dx
\end{equation}
is the onsite interaction energy between atoms with the same spin states.
On the other hand, 
\begin{equation}
\label{eq:U2}
  U_{2} = U_{0} \int \left|w_{0}(x+a/4)\right|^{2}
  \left|w_{0}(x-a/4)\right|^{2}\,dx
\end{equation}
represents the density-density interaction between atoms occupying neighboring sites with 
opposite spin states, i.~e., the interactions acting along the diagonal links of the 
semi-synthetic zigzag lattice shown in Fig.~\ref{fig:setup}(a).
The prefactor $U_0$ is defined by the scattering length (assumed to be state-independent) 
and the confinement in the perpendicular ($y$, and $z$) spatial directions. The specific 
value of $U_0 \approx 1.09\, E_R$ used in our simulations was obtained for the perpendicular 
confinement depths of $30\,E_{R}$. 
In Fig.~\ref{fig:setup}(b), we plot $U_1$ and $U_2$ as a function of the lattice 
depth showing that $U_2$ is around five times smaller than $U_{1}$ for a typical lattice 
height $V_0 = 5\,E_R$. On the other hand, interaction between the atoms at the neighboring 
sites with the same spin state is not included because it is much smaller than both $U_1$ 
and $U_2$.

\subsection{Many-body phases}

\begin{figure*}[ht]
\begin{centering}
\includegraphics[width=\textwidth]{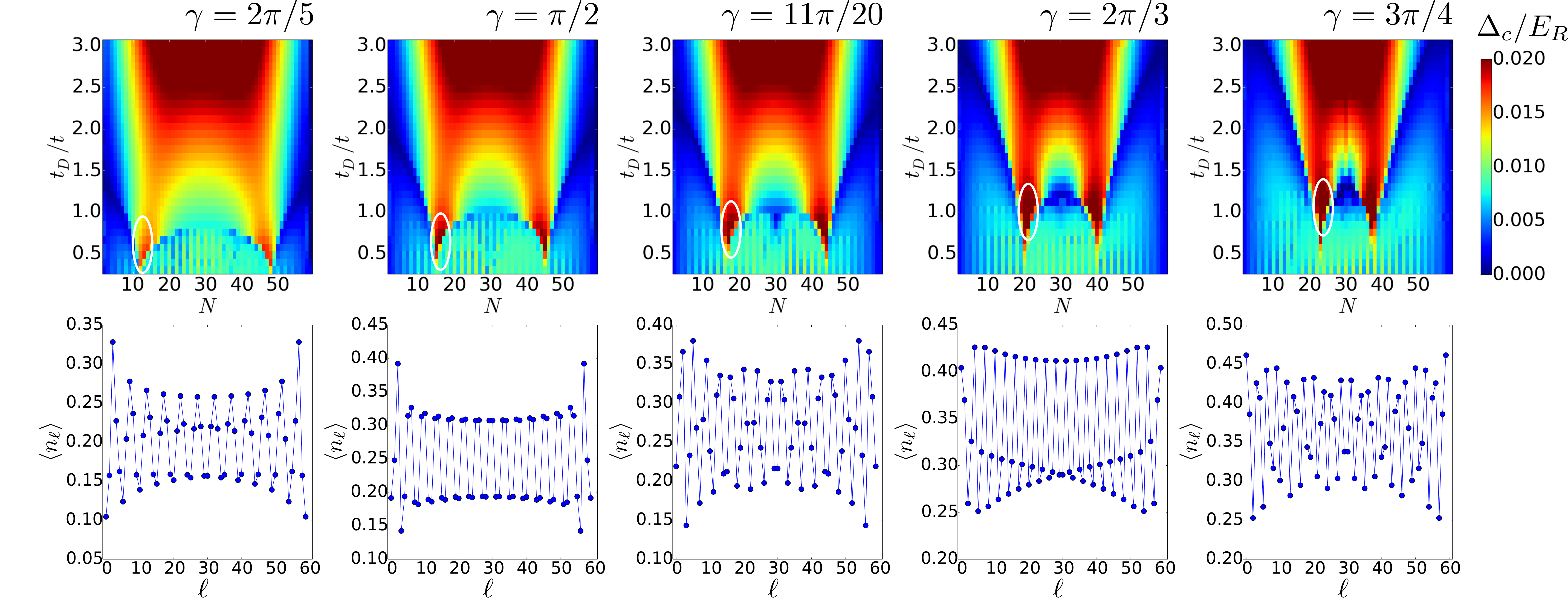} 
\end{centering}
\caption{\label{fig:dwave} Many-body phase diagram of the zigzag lattice for a set of flux 
values $\gamma=\{0.4\,\pi,0.5\,\pi,0.55\,\pi,\tfrac{2}{3}\pi,\tfrac{3}{4}\pi\}$.
Top row: the scaled charge gap $\Delta_{c}/E_{R}$ plotted as a function of the number 
of particles $N$ in the lattice of size $L=60$ and the ratio of the hopping strengths 
$t_{D}/t$. Areas corresponding to enhanced charge gaps close to $t_{D}\approx t$ and 
the filling factor $N/L=\gamma/2\pi$ are conspicuous and are marked with white ovals. 
Bottom row: the expectation of the site occupation $\langle n_{\ell}\rangle$ versus the 
site number $\ell$ calculated at the phase-diagram points inside the white ovals.}
\end{figure*}

In our calculations we take the lattice height $V_{0} = 5\,E_{R}$ for which the interaction 
energies read $U_1 \approx 0.56 \, E_R$ and $U_2 \approx 0.074\, E_R$. To investigate the 
many-body phases supported by the semi-synthetic zigzag lattice we 
performed a series of numerical simulations based on the density-matrix renormalization 
group technique \cite{Schollwock2011} using the open-source OSMPS code \cite{Wall2012}. 
Our simulations targeted the ground states of lattices containing $L = 60$ sites (that is, 
$30$ two-site unit cells) with open boundary conditions and fractional filling factors $N/L$ 
corresponding to all integer particle numbers $N$ up to $60$. Working with such finite 
systems we were able to stay close to the experimentally feasible regime \cite{Atala2014}
while also maintaining a reasonable numerical effort. To check the scaling properties 
of the obtained results representative simulations were rerun also with larger lattice 
sizes containing up to $120$ sites. The remaining two parameters whose values were tuned 
in a broad interval are the flux $\gamma$ and the diagonal hopping parameter $t_{D}$. 
On the other hand, the values for the horizontal hopping parameter $t \approx 0.066 \, E_R$ 
and the nearest-neighbor interaction strength $U_2$ were taken from the modeling 
of a lattice of depth $V_{0} = 5E_{R}$ [cf.\ Fig.~\ref{fig:setup}(b)]. Focusing on the 
effects brought about by the strong atom-atom interactions, in the main part of our 
calculations we chose to work in the limit of hardcore bosons. Thus, the onsite interaction 
strength $U_{1}\approx0.56\,E_{R}$ is regarded to be the dominant energy scale and is accounted 
for by restricting the number of bosons per lattice site to be not more than one. Having rerun 
the calculations with more than one boson per site we were able to confirm that the observed 
interesting many-body phases described below are indeed adequately represented by the hardcore
limit.

The zigzag lattice offers a possibility to realize a tunable single-particle dispersion, 
seen in Fig.~\ref{fig:dispersion}, by changing the ratio of the diagonal and horizontal tunneling 
parameters $t_{D}/t$. In the limits where one of these quantities significantly exceeds 
the other, $t_D \gg t$ or $t \gg t_D$,
we observe quasicondensed phases signaled by the algebraic decay of the 
single-particle density matrix $g_{1}(i,j)\equiv\langle c_{i}^{\dag}c_{j}\rangle$ as a function 
of the separation of sites $|i-j|$ 
\footnote{In performing calculations we represent the zigzag lattice as a one-dimensional 
array of sites enumerated consecutively along a zigzag-shaped path.}. 
In the limit of a dominant diagonal tunneling $t_{D}$, one obtains the usual quasicondensate 
at the single minimum at $k = 0$. Since the magnetic flux is not absorbed into the internal 
structure of the quasicondensate wave function with $k = 0$, the chiral currents are induced 
in the legs of the lattice \cite{Atala2014,Piraud2015}. This phase supported by the zigzag 
lattice corresponds to the one observed in square ladders \cite{Atala2014,Piraud2015}. 
It has been termed the \emph{Meissner phase} in analogy to the physics of superconductors. 
In the opposite limit of weakly coupled spin-polarized legs -- that is, when the horizontal 
hopping $t$ is dominant -- we find a striped phase analogous to the \emph{vortex phase} 
formed in square ladders \cite{Atala2014,Piraud2015}. Here, the current and density 
oscillations are induced by the interference of partial quasicondensates occupying the two minima 
in the single-particle band structure. While this qualitative picture is strictly valid 
for non-interacting bosons it does survive also in the presence of finite interactions. 
Let us also stress that in the thermodynamic limit (as opposed to finite-size simulations) 
the gapless vortex phase is expected to support oscillations in the density correlations 
and not the density itself.

In between the two limits supporting quasicondensed ground states there lies an 
intriguing regime of balanced tunneling strengths $t \sim t_{D}$ associated with the 
presence of kinetic frustration. In particular, when $t = t_{D}$ each triangular plaquette 
is characterized by the absence of a weak link that could absorb the complex phase 
accumulated while encircling the plaquette. Under such circumstances the role of the 
atom-atom interactions will be enhanced, which might drive the system into a gapped phase. 
Indeed, our simulations show that the power-law decay of the single-particle density matrix 
$g_{1}(i,j)$ is replaced by an exponential decay signaling the destruction of the quasicondensed 
phase. To complement these observations, in Fig.~\ref{fig:dwave} we plot the behavior of the 
charge gap \cite{Rossini2012} $\Delta_{c}(N) = E_{N+1} + E_{N-1} - 2E_{N}$ calculated from 
the ground-state energies of the zigzag lattice with a varying number of particles. 
In the top row, the coordinate axes represent the two governing parameters, the filling 
factor $N/L$ (with $L = 60$) plotted on the horizontal axis and the ratio of the hopping 
parameters $t_{D}/t$ plotted on the vertical axis. The series of five phase diagrams represent 
a subset of calculations performed on a dense set of different values of the flux $\gamma$. 

The phase diagrams reveal the emergence of areas -- marked with white ovals -- where 
charge gaps are significantly enhanced. It is noteworthy that these gapped ``islands'' 
are situated precisely at the parameter values where the single-particle correlations 
decay exponentially and the filling factor assumes flux-dependent values $N/L = \gamma/2\pi$ 
and $N/L = 1 - \gamma/2\pi$. These two values are related by the particle-hole symmetry 
brought about by the hardcore constraint. They correspond precisely to the situation with 
one particle or hole per magnetic unit cell containing $2\pi/\gamma$ triangular lattice 
plaquettes or $2\pi/\gamma$ sites, like in the integer bosonic Hall effect 
\cite{Senthil2013,Furukawa2013,Wu2013,He2015}.

The bottom row of panels in Fig.~\ref{fig:dwave} shows the particle density oscillations 
calculated at points taken from inside the white ovals. Here, the expectation value of 
the density $\langle n_{\ell}\rangle$ is plotted as a function of the site index
$0\leqslant\ell\leqslant59$. We see that for any value of the scaled flux $\gamma$ and 
the corresponding flux-dependent filling $N/L = \gamma/2\pi$, the density oscillations 
occur at the wavelength corresponding to one particle per oscillation. For example, in the 
second column we look at $\gamma = \pi/2$ and the filling $N/L = 1/4$, thus implying 
$N = 15$ for $L = 60$. Here we count $15$ full oscillations of the density, each covering 
four sites. Note that the observed density wave is fundamentally different from a gapped 
phase with staggered density modulation, which is directly favored by strong nearest-neighbor 
interactions 
$U_2$ and found at half filling (see the following subsection), 
since it occurs on longer wavelengths dictated 
by the magnetic flux. Nevertheless, a finite value of $U_2$ enhances the charge gap of the 
flux-induced density wave. As seen in the plots corresponding to $\gamma = 2\pi/5$ and 
$\gamma = 2\pi/3$, periodicities of three and five sites are also possible. The remaining 
two panels are calculated at $\gamma = 11 \pi / 20$ and $\gamma = 3\pi/4$. Here, according 
to the general observed trend one expects an incommensurate filling of, respectively, 
$16.5$ and $22.5$ particles per $60$ sites. Although the density distributions look less 
regular in these cases, one still observes the formation of a density wave following the 
same predictive pattern. The required filling corresponds to the density where the magnetic 
unit length matches the wavelength of Friedel oscillations \cite{Friedel58} in a system of 
free fermions, to which the simulated system can be mapped for $t = U_2 = 0$. 
Friedel oscillations occur near local defects, such as the boundary of the system, 
and decay algebraically. One can see in Fig.~\ref{fig:dwave} that (at finite $U_2$) 
they are promoted to a long-ranged density wave and persist across the sample, when 
a commensurate magnetic length is introduced with a finite value of $t$.

\subsection{Spin polarization}

\begin{figure}
\begin{centering}
\includegraphics[width=80mm]{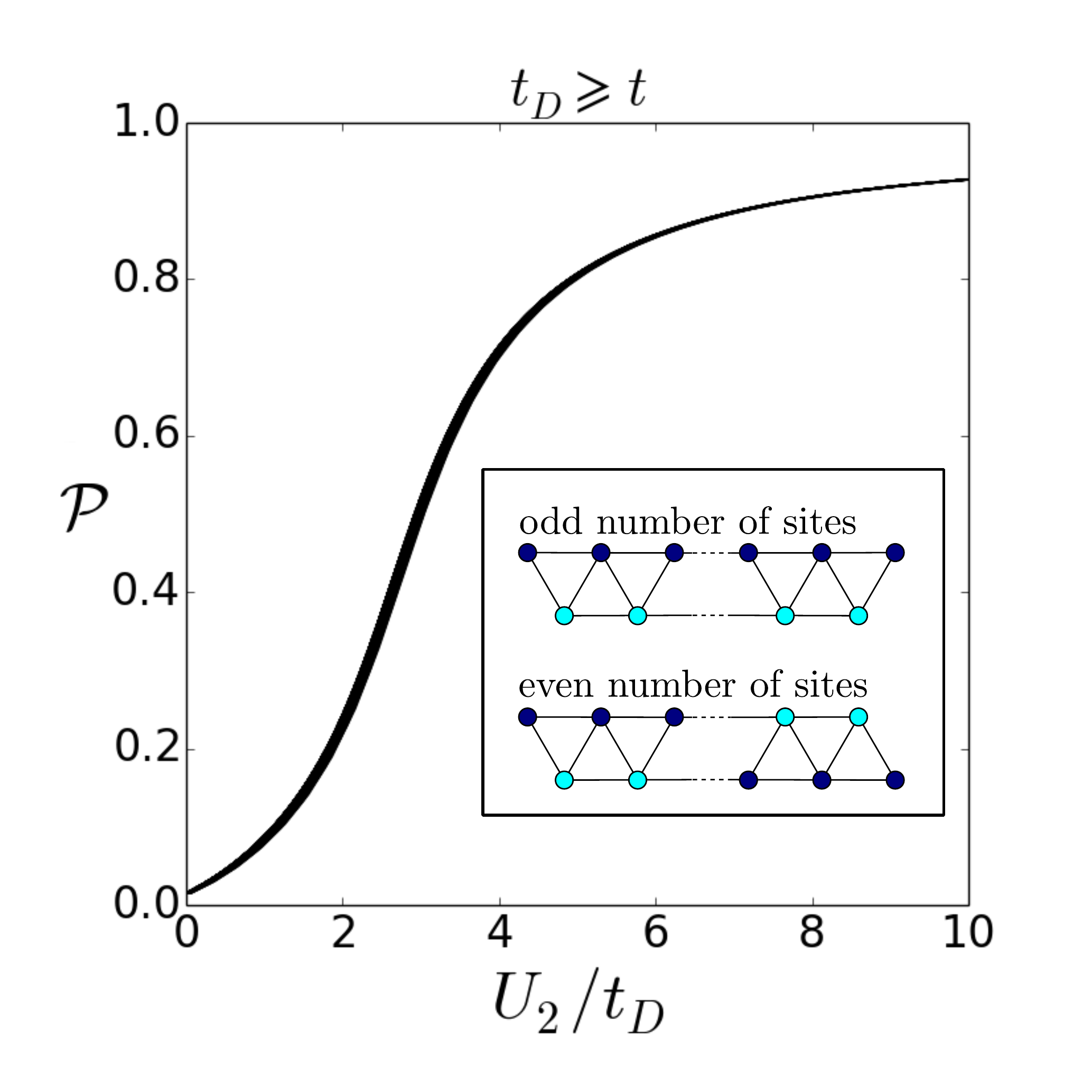} 
\par\end{centering}
\caption{\label{fig:polarization} The spin polarization $\mathcal{P}$ as a function
of the nearest-neighbor repulsion strength calculated for $N=30$
particles on a zigzag lattice of $L=59$ sites without the artificial
flux. When diagonal transitions dominate over the horizontal transitions,
$t_{D}\geqslant t$ the shown dependence of the polarization $\mathcal{P}$
becomes insensitive to the relative magnitude of $t$.
The thick line is a superposition of many dependencies with the ratio of 
the hopping parameters $1\leqslant t_{D}/t\leqslant20$.}
\end{figure}

As a further example of the many-body physics supported by the zigzag lattice we look at 
the spin polarization induced by the strong nearest-neighbor interactions. Here, we work 
at the average filling $N/L$ close to one half and in the absence of the artificial flux, 
$\gamma=0$. In the presence of nearest-neighbor interaction \textemdash\ which is
a distinguishing feature of the zigzag configuration \textemdash\ the particles are 
expected to occupy every second site thereby preferentially flocking onto one of the 
ladder legs and inducing non-zero net spin polarization defined as 
$\mathcal{P} = (n_{\uparrow}-n_{\downarrow})/(n_{\uparrow}+n_{\downarrow})$.
A numerical calculation reveals that in this particular case the supported ground state 
configuration is sensitive to the total number of sites being even or odd. The effect can 
be explained in a simple way as the tendency of strong interactions to push particles into 
the sharp corners formed at the ends of the finite lattice. This is illustrated in the inset 
of Fig.~\ref{fig:polarization} where the dark (light) blue color is used to mark preferentially 
occupied (depleted) lattice sites. Obviously, if the total number of sites $L$ is even, 
the boundary conditions lead to opposite preferred spin polarizations at the two ends of 
the finite lattice, and the polarization must change sign somewhere in the middle of the 
lattice. In contrast, when the total number of sites is odd, the boundary conditions 
facilitate the largely uniform spin polarization of the whole lattice. 

As a matter of fact,
whether or not such a spin-polarized configuration will be formed depends on the competition 
of the nearest-neighbor repulsion and the delocalizing effect of inter-site hopping. The 
results of our numerical simulations performed on the system of $N=30$ hardcore particles
on a lattice of $L=59$ sites are presented in Fig.~\ref{fig:polarization}. It is striking that as 
soon as $t_{D}\geqslant t$ the behavior of the spin polarization $\mathcal{P}$ shows a 
universal behavior -- it depends only on the the ratio $U_{2}/t_{D}$ and is virtually 
independent of the strength of the relatively weaker spin-preserving transitions with the 
parameter $t\leqslant t_{D}$. The thick line shown in Fig.~\ref{fig:polarization} is in fact a 
superposition of many dependencies with the ratio of the hopping parameters 
$1\leqslant t_{D}/t\leqslant20$. In the complementary regime $t\geqslant t_{D}$, 
spin-preserving hopping transitions start to contribute to the melting of the spin-polarized 
state. Here, relatively stronger interactions are needed to induce the spin imbalance, 
and the polarization $\mathcal{P}$ depends on both $U_{2}/t$ and $U_{2}/t_{D}$ 
and thus loses its universal behavior.

\section{Summary}\label{sec:summary}

We proposed a scheme for the realization of
a semi-synthetic zigzag optical lattice built from a one-dimensional 
spin-dependent optical lattice with transitions between internal atomic states. Each 
of the lattice's triangular plaquettes ensnares the same---tunable---magnetic flux that 
can controllably deform the single-particle band structure form the single-minimum to the 
double-well configuration. In the proposed setup, the atom-atom interactions are nonlocal 
in both dimensions and stabilize density-wave-like phases at flux-dependent filling factors.

We thank 
Immanuel Bloch, 
Alessio Celi, 
Xiaoling Cui,
Simon F\"{o}lling, 
Sebastian Greschner,
Maciej Lewenstein, 
Michael Lohse, 
Pietro Massignan,
Leonardo Mazza,
Shuyan Wu,
and 
Jakub Zakrzewski 
for helpful discussions. This research was supported by the Lithuanian Research 
Council (Grant No.\ MIP-086/2015) and by the German Research Foundation (DFG) via the 
Research Unit FOR 2414. I.~B.~S.\ was partially supported by the ARO's Atomtronics MURI,
by AFOSR's Quantum Matter MURI, NIST, and the NSF through the PCF at the JQI.
C.~S.\ is grateful for support by the Studienstiftung des deutschen Volkes. 

\bibliography{zigzag}

\end{document}